# A Jamming-Resilient MAC-layer Device Identification for Internet of Things*

Przemysław Błaśkiewicz, Jacek Cichoń, Jakub Lemiesz,
Mirosław Kutyłowski, Marcin Zawada
Wroclaw University of Technology, Wroclaw, Poland
Email: {firstname.lastname}@pwr.edu.pl

Krystyna Napierała, Michał Panek,
Stanisław Strzyż
DATAX Sp. z o.o., Wroclaw
Email: knapierala@datax.pl, mpanek@datax.pl,
sstrzyz@datax.pl

## I. Introduction

In a number of practical scenarios a wireless device needs to mark its presence, for instance, to some access point. That enables the access point to assign the device its transmission slot or update the count of the network nodes. Many protocols can achieve exactly this result. In this paper, our goal is to show how that can be done in the simplest messaging model, the so-called beeping model. Consequently, we constrain our design so that the station does not send any modulated information in the signal and the receiver actually does not need to demodulate/decode it. We are interested in sending just a short signal, so called **beep**. Moreover, we want to design such protocol that is resilient to random interference and enables us to identify devices which are sending the signal, as opposed to only note their presence. To do that, we leverage temporal correlations of a sequence of beeps issued by a device, as if the time-moments when they happen come from a pre-defined probability distribution, that is the fingerpring of the device.

## II. Related work

The beeping model was first introduced in [1], where the interval coloring problem for a wireless network was shown suitable under this model constraints. Following work of Cornejo and Kuhn, in [2], the authors presented an algorithm for calculating maximal independent set under the same assumptions. In [3] the same model was applied for spontaneous, asynchroneous and secure communication under strong adversary assumptions.

The above papers are, to our knowledge, the only ones utilising the carrier sensing (CS) mechanism directly for reception of messages. Some influence of CS on standard protocols in wireless communication in described in [4]. The main application of CS is that of creating MAC (multiple access channel) protocols [5], [6]. In [7], [8], the carrier sensing was employed for network initialization. The same mechanism can facilitate fast construction of dominating sets in sensor networks [9].

*Supported by National Centre for Research and Development project no. PBS1/A3/6/2012

## III. Model

### A. Beeping model

The model of our interest is arguably the most constrained communication model in the literature. We assume that a station can communicate in one way only, that is by sending a *beep*. In other words, it can either alter the transmission medium (beeping), or not (silence). More, the station transmits regardless of other transmissions, i.e. there is no collision avoidance mechanism. The receiving site discovers the beep by CS (carrier sensing) mechanism, and it is impossible to distinguish between cases when a single station was sending its beep or more stations were transmitting at the same moment and the beeps colided.

Realistically, one must put a time duration for a single beep, and that might be the time needed for CS hardware to pick up the signal under given conditions of transmission power and distance. Regardless of the method, we assume that a beep has a pre-determined duration, and use this value as a basic interval for any operations performed in the algorithms. Note however that this does not mean a slotted model – indeed, while there is a fixed underlying time period for all activities, they can happen at any moment, so consequently we assume an asynchronous model of messaging.

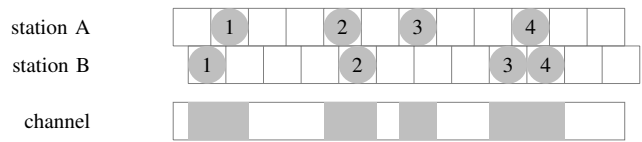

Fig. 1. Channel representation in case of two stations transmitting. Note how first and second beeps of 'A' and 'B' get glued together, the third beep of 'A' gets through in clear, but its last beep is totally obscured by the two consecutive beeps from 'B'.

## IV. New algorithm

In this section we assume that we have $n$ devices that want to show their presence in the system. Let us assume that each device has it own *pattern*, representing points in time, at which the device will be beeping. Thus the access point will only receive the union of all beeps and on that basis it should recognize the devices ids.

**Algorithm 1** BERNOULIIBROADCAST($id, p$)

**Initialization:**
1: rand seed ← $id$

**At time $t \in [1, T]$:**
1: **if** rand$(0, 1) \leq p$ **then**
2:     broadcast beep at time $t$
3: **end if**

**Identification of the devices:**
1: let $U$ represent table of all beeps received
2: $S \leftarrow \emptyset$
3: **for** $id \in$ ID **do**
4:     rand seed ← $id$
5:     state ← $true$
6:     **for** $t = 1$ to $T$ **do**
7:         **if** rand$(0,1) \leq p$ and $U[t] \neq 1$ **then**
8:             state ← $false$
9:         **end if**
10:     **end for**
11:     **if** state = $true$ **then**
12:         $S = S \cup \{id\}$
13:     **end if**
14: **end for**

*A. Analysis*

At first, we calculate the false identification of the algorithm 1. It means that station with identifier $id$ is recognize by our algorithm, but it didn't send beeps signal. Assume that $n$ is number of stations and $k$ denotes the number of beeps signal.

**Lemma 1.** *The probability, that in algorithm* BERNOULIIBROADCAST($id, p$) *each $n$ of $n+1$ stations choose exactly the same $k$ beeps is equal to* $(1-(1-p)^n)^k$.

*Proof:* Notice that stations that transmit beeps signal follow the Bernoulli distribution. Let $X$ be random variable from Bernoulli distribution with probability of success $p$. Then the probability that at least one station transmitted the beep is given by

$$P(X \geqslant 1) = 1 - (1-p)^n.$$

Now, we need to identify if at each of $k$ slots we had a transmission. Thus we obtain

$$(1-(1-p)^n)^k.$$

■

However in our algorithm the number of slots is also random variable that follow the Bernoulli distribution.

**Theorem 1.** *The probability that the station is falsely identify by algorithm 1 is given by*

$$(1 - p(1-p)^n)^T. \quad (1)$$

*Proof:* Let $Y_i$ be the Bernoulli $\mathcal{B}(T, p)$ random variable that represent the number of slots at which $i$-th station has transmitted. By lemma 1 we get the probability of false identification

$$\sum_{k=0}^{n} (1-(1-p)^n)^k P(Y_i = k) .$$

Therefore

$$\sum_{k=0}^{m} (1-(1-p)^n)^k \binom{m}{k} p^k (1-p)^{n-k} = (1 - p(1-p)^n)^T .$$

■

Using equation (1) we can minimize the probability of false identification by finding the optimal transmitting probability $p$. Calculating the derivative and after simplification we obtain

$$-(1-p)^n + n(1-p)^{n-1} p = 0 .$$

Solving the above equation we get

$$p_{opt} = \frac{1}{n+1} .$$

Then the optimal probability (1) is equal to

$$\left(1 - \frac{\left(1 - \frac{1}{n+1}\right)^n}{n+1}\right)^T \approx \left(1 - \frac{1/e}{(n+1)}\right)^T .$$

Our goal is the minimization that algorithm can identify falsely some station, therefore we calculate

$$\left(1 - \frac{1/e}{(n+1)}\right)^T = \frac{1}{n} \quad (2)$$

Then we can find the optimal parameter $T$ (see pseudocode) as follows:

$$T = \frac{\ln(1/n)}{\ln(1 - \frac{1}{e(n+1)})} .$$

V. SIMULATION ASSUMPTIONS

To check the performance of proposed algorithm, a series of Monte Carlo simulations has been performed. Summary of parameters was given in Table I.

TABLE I
SIMULATION PARAMETERS SUMMARY

| Simulation runs | 50 |
|---|---|
| Simulation length | 5s |
| $t$ | 10ms |
| $T$ | [50, 100, 150, 200, 500, 1000] ms |
| $n$ (number of nodes) | 10 (5 active) |
| $p$ | [0.1, 0.2, 0.3, 0.4, 0.5] |
| Interference rate | [0, 0.01, 0.02, 0.05, 0.1, 0.2] |
| Node TX power | $10\mu W$ |
| Receiver sensitivity | -104 dBm |
| Std. deviation of shadow fading | 8dB |
| Fading Type | Rayleigh fading |
| Node velocity | 3 kmph |

Simulations were modelling an environment with 10 radio nodes, out of which 5 were transmitting it's patterns. One central node was receiving transmissions and every $T$ was performing an identification of nodes. Nodes were dropped randomly in a square, 100m x 100m area, with receiving node in a central position. Each node was transmitting with constant power equal to $10\mu W$, while the receiver sensitivity was -104 dBm. Rayleigh fading was modelled resulting in dynamic changes in radio conditions. Each transmitting node was assigned a velocity, which was introduced to model fast fades properly, but didn't result in any position changes. Additionally, external random interference was introduced, modeled by random transmissions of strong, external signal. The fraction of time when the external interference is present to the whole simulation time was described by *Interference rate*. For each parameter set, 50 simulations has been performed, while each run modelled 5 seconds of network operation.

## VI. SIMULATION RESULTS

In this section the results of simulations are presented along with analysis thereof. On the figures, two types of results are presented, namely probability of identification of transmitting node (True Positives Rate, denoted as $TP\ Rate$ hereafter) and probability of correct classification of silent node (True Negative Probability, denoted as $TN\ Rate$ hereafter). As an example, consider a situation, where a node, that transmitted its signal, was identified by a listening node. This situation is marked as True Positive. On the other hand, if a node that remain silent, was not identified as transmitting one in the receiving node, this situation is considered True Negative. Other options, which are complementary to the abovementioned, are False Positives, when the silent node was identified as transmitting by the receiving node, and False Negatives, when the active node was not identified by the receiving node. The distinction between True Positives and True Negatives is important, as the algorithm should not only identify correctly nodes which are transmitting, but should avoid false alarms. Typically, there is a tradeof between those two values.

### A. Ideal conditions

First set of results was assuming no attenuation of radio signal. The purpose was to understand the behavior of the algorithm in the perfect conditions and to distinquish between deteriorations of algorithm's performance coming from radio conditions and from the algorithm itself. Obviously, when radio signal from transmitting nodes is neither attenuated nor interfered by external sources of signal, the TP rate is constantly equal to 100%. What is interesting, is the TN rate depicted in Fig. 2.

The TN rate is increasing with the increase of $T$ and is decreasing with the increase of $p$. The first effect can be explained by the fact, that the longer $T$ is, the better is the ability of the algorithm to produce unique beep sequences, assuming constant number of nodes. The impact of $p$ is opposite, higher values results in radio environment being more congested (nodes are transmitting more frequently), leading to lower ability of the receiver to correctly classify nodes that are silent.

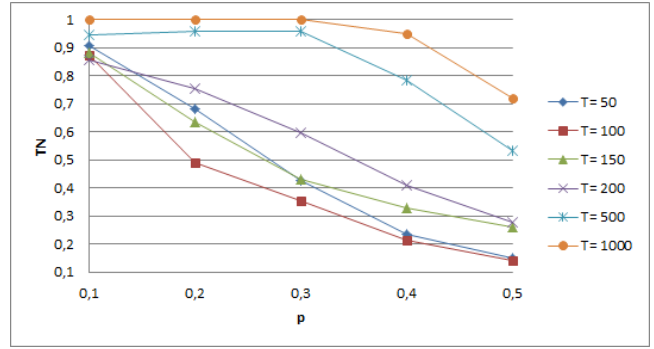

Fig. 2. TN rate vs $p$, for variety of $T$ values, ideal radio conditions

### B. Normal radio conditions

In this subsection results are presented in which realistic radio conditions were taken into account. First, simulation results with no external interference are presented. In those experiments, each transmission is a subject to fast fading, shadow fading and pathloss, which models isolated environment.

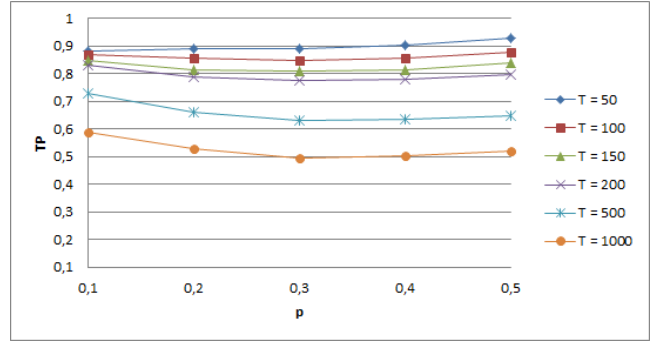

Fig. 3. TP rate vs $p$, $IR = 0$, for variety of $T$ values

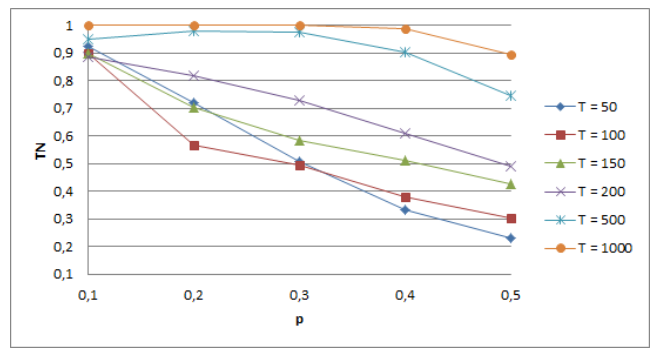

Fig. 4. TN rate vs $p$, $IR = 0$, for variety of $T$ values

As can be seen in figure 3, TP rate increases with the length of $T$. The influence of $p$ parameter can be seen mostly in case of higher values of $T$, but generally the TP level is rather insensitive to the changes of $p$. As mentioned above, in ideal radio conditions the TP rate is constant ad equal to 1. Here, the reason for a decrease of TP rate is the impact of fast fading. For short time, signal transmitted from a given node, might experience massive attenuation, leading to receiver's inability to sense the transmitted signal, as the received power is below receiver's sensitivity.

For the TN rate (Fig. 4), the results are analogous to those without attenuation (see Fig. 2). A decrease with the increase of $p$ is visible, especially for lower values of $T$. The decrease of TN is also dependent on the $T$ value.

In figures 5 and 6 an example is presented, describing the impact of different levels of external interference on TP rate. It can be noted, that the results are only slightly influenced by external interference, leading to small increase of TP rate with increase of $Interference\ Rate$ (up to 2% for the case of $IR$ equal to 0.2). The TN rate is more influenced by outer interference, where an increase of $IR$ leads to decrease of TN rate up to around 8% (for the case of $IR$ equal to 0.2). Similar conclusions can be drawn for other values of $T$, not presented here.

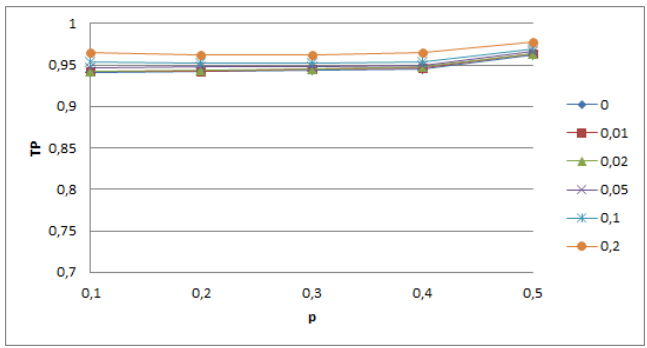

Fig. 5. TP rate vs $p$, $T$ = 100ms, for variety of interference rate ($IR$) values

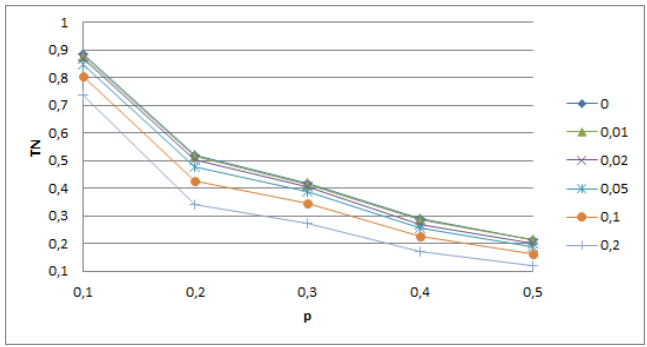

Fig. 6. TN rate vs $p$, $T$ = 100ms, for variety of interference rate ($IR$) values

## C. Filtering

To increase the efficiency of the algorithm in identyfing transmitting nodes, a filtering function was added in the receiver. The main reason for introduction of this enhancement, as mentioned in the previous subsection, is the decrease of TP rate due to fast fadings of transitted signal. It was assumed that the nodes are sending their sequence in several consecutive periods $T$. The receiver, after each beeping period $T$, had statistics about slots in which transmission occured. After $m$ such periods (where $m$ is a filter length), receiver perfomed filtering by performing logical $OR$ function on respective slots of each beeping period. In other words, the algorithm assumed that in given slot the transmission occured, if in any of the beeping periods covered by filtering window, in respective slot transmission was detected. In figure 7 an example of filetring is depicted. Four periods of algorithm are presented, as seen by the receiving node, where each dashed block represents moments, when transmission was detected. As can be seen, it is sufficient if in only one period some transmission gets detected. This operation, on one hand, leads to higher TP rate

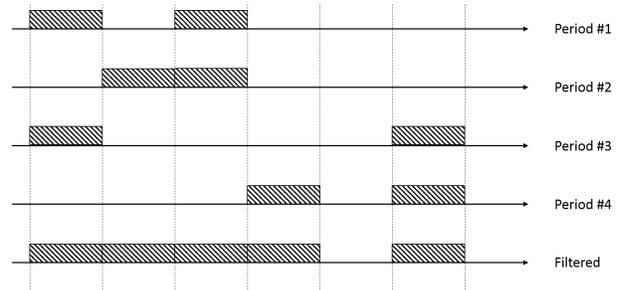

Fig. 7. Filtering

(as the algorithm is no longer impacted by a single fade), but might also lead to increase of TN rate in an environment, where the external interference level is high. In figures 8 and 9 the gains of TP rate and losses of TN rate due to filtering were presented. The worst case scenario was taken into account, where the level of external interference was high.

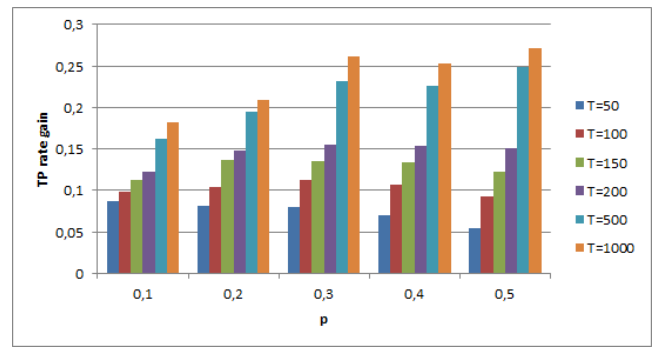

Fig. 8. TP rate gains due to filtering, $Interference\ rate$ = 0.2, filter length = 6

It can be observerd, that filtering lead to increase of TP rate gains of up to 27%, while, due to unfavourable radio conditions ($IR$=0.2), TN rate in some cases was declined by 28%. The following figure explains in which cases filtering brought an increase in overall performance. It presents the difference between gains of TP rate due to filtering and losses of TN rate introduced by filtering, for the respective cases. TP rate gains, as well as TN rate losses were calculated as a difference between results before and after introduction of filtering.

Figure 10 gives an indication on which cases benefit from filtering. It can be seen, that for low values of $T$, the TN

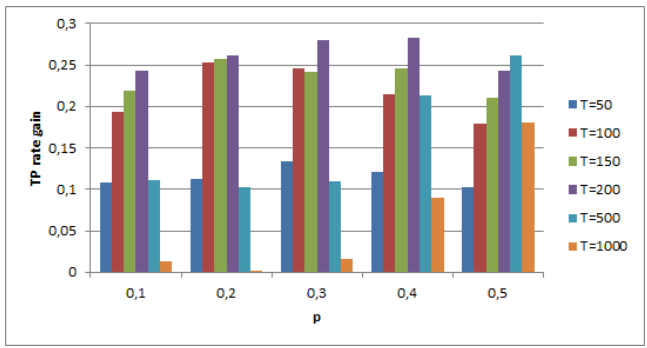

Fig. 9. TN rate losses due to filtering, $Interference\ rate$ = 0.2, filter length = 6

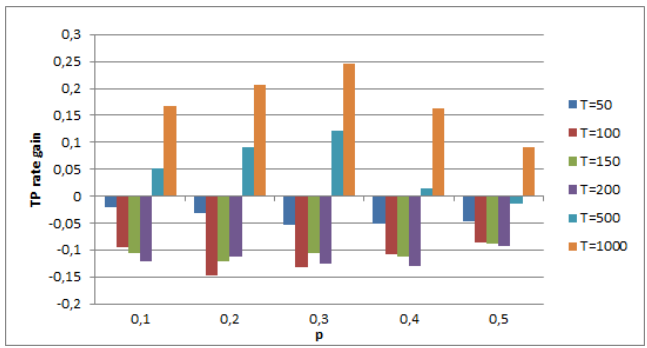

Fig. 10. TP rate gain - TN rate loss, $Interference\ rate$ = 0.2, filter length = 6

rate losses are higher than the TP rate gains, irrespective of the value of $p$. For long beeping periods ($T$=500, 1000ms), significant improvement of overall performance is observed, reaching almost 25%.

## VII. SIMULATIONS - CONCLUSIONS

In this paper, the results of simulations of beeping algorithm with various parameter settings were presented. The results were evaluated based on TP rate (True Positives - correct identification of transitting node) and TN rate (True Negatives, correct classification of silent node). First, the result obtained in simulations assuming ideal radio conditions were performed. TP rate was constant and equal to 1, while TN rate was decreasing with the increase of $p$ and increasing with the increase of $T$. Next, the results assuming realistic radio conditions were performed. Results suggested, that TP rate decreases with the increase of $T$ and is only slightly dependent on $p$. TN rate, on the other hand, increases with $T$, but is inversely proportional to $p$. When the external interference is introduced, it caused a moderate decline in TN rate and insignificantly increased TP rate. To counteract the degradation of TP rate due to fast fading of radio channel between given transmitter and receiver, a filtering function was introduced. Filtering caused an increase in the TP rate, but simultaneously a decline in TN rate in scenarios with high level of external interference. It was noted, that for high values of $T$, the overall performance of the algorithm was improved by the filtering, meaning that TP rate gains were significantly higher than the losses of TN rate due to filtering. It has to be emphasized, that the lower the external interference, the more gains the filtering brings.

## VIII. FUTURE IMPROVEMENTS

Despite performing extensive studies on the subject, authors see a possible improvements of the algorithm. One such enhancement could be e.g. designing pseudorandom generator in such a way, that for given set of parameters and some roughly estimated or upper-limited number of nodes in the network, it produces sequences, which Hamming distance is maximized. In such a case it is possible to improve the performance of the algorithm, both in terms of TP rate and TN rate, by further minimization of the influence of fast fadings and random interference.